\documentclass[letterpaper,twocolumn,onecolappendix,appendixfloats]{emulateapj}
\usepackage{float}
\usepackage{epsfig}
\usepackage{graphicx}
\usepackage{graphics}
\usepackage[latin1]{inputenc}
\usepackage{latexsym}

\shorttitle{Photometry of AST3-1 from Dome A}
\shortauthors{Wang et al.}

\begin{document}

\title{Variable stars observed in the Galactic disk by AST3-1 from Dome A, Antarctica}

\author{Lingzhi Wang\altaffilmark{1,2,3}, Bin Ma\altaffilmark{1,3}, Gang Li\altaffilmark{4}, Yi Hu\altaffilmark{1,3}, Jianning Fu\altaffilmark{4}, Lifan Wang\altaffilmark{3,5,6}, Michael C.~B.~Ashley\altaffilmark{7}, \\Xiangqun Cui\altaffilmark{3,8}, Fujia Du\altaffilmark{3,8}, Xuefei Gong\altaffilmark{3,8}, Xiaoyan Li\altaffilmark{3,8}, Zhengyang Li\altaffilmark{3,8}, Qiang Liu\altaffilmark{1,3}, Carl R.~Pennypacker\altaffilmark{9}, \\Zhaohui
  Shang\altaffilmark{1,3,10}, Xiangyan
  Yuan\altaffilmark{3,8}, Donald G.~York\altaffilmark{11}, Jilin Zhou\altaffilmark{3,12}}

\altaffiltext{1}{National Astronomical Observatories, Chinese Academy of Sciences, Beijing 100012, China. wanglingzhi@bao.ac.cn
}

\altaffiltext{2}{Chinese Academy of Sciences South America Center for Astronomy, Camino EL Observatorio 1515, Las Condes, Santiago, Chile}

\altaffiltext{3}{Chinese Center for Antarctic Astronomy, Nanjing 210008,
  China}

\altaffiltext{4}{Department of Astronomy, Beijing Normal University, Beijing,
  100875, China}
\altaffiltext{5}{Purple Mountain Observatory, Chinese Academy of
  Sciences, Nanjing 210008, China}

\altaffiltext{6}{Mitchell Institute for Fundamental Physics \& Astronomy,
  Department of Physics \& Astronomy, Texas A\&M University, College Station,
  TX 77843, USA}

\altaffiltext{7}{School of Physics, University of New South Wales, NSW 2052,
  Australia}

\altaffiltext{8}{Nanjing Institute of Astronomical Optics and Technology,
  Nanjing 210042, China}

\altaffiltext{9}{Center for Astrophysics, Lawrence Berkeley National
  Laboratory, Berkeley, CA, USA}

\altaffiltext{10}{Tianjin Normal University, Tianjin 300074, China}

\altaffiltext{11}{Department of Astronomy and Astrophysics and Enrico Fermi
  Institute, University of Chicago, Chicago, IL 60637, USA}

\altaffiltext{12}{School of Astronomy and Space Science and Key Laboratory of Modern Astronomy and Astrophysics in Ministry of Education, Nanjing University, Nanjing 210093, China}

\begin{abstract}
  AST3-1 is the second-generation wide-field optical photometric telescope dedicated to time domain astronomy at Dome A, Antarctica. Here we present the results of $i$ band images survey from AST3-1 towards one Galactic disk field. Based on time-series photometry of 92,583 stars, 560 variable stars were detected with $i$ magnitude $\leq$ 16.5 mag during eight days of observations; 339 of these are previously unknown variables. We tentatively classify the 560 variables as 285 eclipsing binaries (EW, EB, EA), 27 pulsating variable stars ($\delta$~Scuti, $\gamma$~Doradus, $\delta$~Cephei variable and RR Lyrae stars) and 248 other types of variables (unclassified periodic, multi-periodic and aperiodic variable stars). Among the eclipsing binaries, 34 show O'Connell effects.  One of the aperiodic variables shows a plateau light curve and another one shows a secondary maximum after peak brightness. We also detected a complex binary system with RS CVn-like light curve morphology; this object is being followed-up spectroscopically using the Gemini South telescope. 

\end{abstract}


\section{Introduction}
Time domain astronomy is the investigation of astronomical objects as a function of time, and has long been a source of interesting and unexpected discoveries. On-going and new ground- and space-based large synoptic sky surveys, such as the (intermediate) Palomar Transient Factory \citep{Law2009,Rau2009}, the SkyMapper Telescope \citep{Keller2007}, and the Large Synoptic Survey Telescope \citep{LSST} after its first light in 2020 \footnote{https://www.lsst.org/about/timeline}, are exploring or will explore new regions of parameter space in terms of depth and temporal coverage.

The Antarctic plateau offers a number of unique advantages for precision, ground-based, time-domain astronomy, such as the ability to observe continuously during winter, low scintillation noise, excellent seeing above a very low boundary layer, low airmass variations, low aerosols, low water vapor, more stable atmospheric transmission, wider wavelength windows, and a dark sky in the infrared \citep{Lawrence04,Lawrence06,Lawrence08,Moore08,Kulesa08,Aristidi09,Burton10,Zou10,Yang10,Sims10,Bonner10,Lascaux11,Tremblin11,Pei11,Pei12,Sims12a,Sims12b,Giordano12,Storey13,Hu14,Ashley13,Yang16}. There is thus considerable interest in overcoming the technical challenges of operating in Antarctica, so that the advantages for astronomy can be realized \citep{Tothill08,Kulesa08,Crouzet10,Chapellier16,Mekarnia16}.

Dome A (latitude $80^{\circ}22'02''$S, longitude $77^{\circ}21'11''$E, elevation 4093m above the sea level) is the highest region on the Antarctic plateau and is being used for a series of three increasingly ambitious optical survey telescopes \citep{Yang09,Gong10}. The first optical telescope was called CSTAR \citep[the Chinese Small Telescope ARray;][]{Yuan08} with an effective aperture of 10\,cm and field of view (FOV) of 20\,deg$^2$ and was installed at Dome A in January 2008; CSTAR produced a 3-year photometric dataset, and a number of studies of stellar variability have been published \citep{Zhou10a,Zhou10b,Wang11,Wang12,Zhou13,Wang13a,Wang13b,Meng13,Huang13,Fu14,Qian14,Wang14a,Wang14b,Zong15,Huang15,Wang15,Yang15,Oelkers15,Oelkers16,Liang16}. The second-generation of optical telescope at Dome A is called AST3, which in turn consists of three telescopes, each with an entrance pupil diameter of 0.5\,m, and a FOV of 4.3\,deg$^2$ \citep[][]{Cui08,Yuan12,Yuan14,Yuan15}. The first two of the AST3 telescopes---AST3-1 \citep{Lix12,Liz12,Wen12} and AST3-2---were installed at Dome A in January 2012 and 2015, respectively. The third AST3 telescope is planned to be installed in 2017, and will have a K-band infrared camera. The third-generation optical/infrared telescope destined for Dome A is called KDUST \citep[the Kunlun Dark Universe Survey Telescope;][]{Jia12,Jia13,Yuan13,Zhu14,Burton16,Li16,Xu16}, which has an aperture of 2.5\,m and FOV of $\sim$2.3\,deg$^2$ \citep{Yuan13}; KDUST is expected to be operational at Dome A after 2022.

The three AST3 telescopes were originally conceived as multi-band survey telescopes, with each telescope having a fixed filter to reduce the risk of mechanism failure. Their main sky survey area is a zenith distance less than 70$^{\circ}$ \citep{Yuan12}. Meanwhile, other factors for observations are also taken into account, including the altitude and phase of the Moon, the angular distance between the telescope pointing and the Moon, and the altitude of the Sun \citep{Shang12}. To operate the AST3 at remote Dome A, an improved version of PLATO (an automated observatory platform for CSTAR and other earlier instruments \footnote{http://mcba11.phys.unsw.edu.au/$\sim$plato-a}), PLATO-A was designed to offer about 1\,kW power source for AST3 \citep{Lawrence09,Ashley10,Shang12}.

The three main science goals for AST3 are the early detection of supernovae, exo-planet transit searches, and stellar variability\citep{Cui08}. The first AST3 telescope (AST3-1) was deployed to Dome A in January 2012 successfully and $\sim$\,16,000 scientific frames were collected from March 16 to May 7, 2012, with a total exposure time of 189 hours. After that, AST3-1 unfortunately stopped work due to a malfunction a power distribution box. Of the $\sim$\,16,000 images obtained, 4,000 were of 500 fields mainly surveyed for supernova templates; $\sim$\,4700 images were of the center of the Large Magellanic Cloud, and $\sim$\,3400 images covered eight Galactic disk fields to study Wolf-Rayet stars, and one Galactic disk field was used primarily to search for transiting exo-planets. This latter field had the most number of observations, and so was also suited to a study of stellar variability, which is the subject of this paper. The field was centered at $l=289.6347^{\circ}$, $b= -1.5718^{\circ}$, and was monitored in $i$ band with 3523 images over 8 days with a total exposure time of 38.9 hours. Of these 38.9 hours, 157 frames totalling 2.6 hours were observed on March 28,  and 3366 frames totalling 36.3 hours were from April 24 to May 1 (Table~\ref{tb:log}). The distribution of the 36.3 hours over the 7 days of observations can be seen from the time-series plots. Gaps in the observations were mainly due to the AST3-1 telescope being pointed to other fields. The observations, the data reduction, and the time-series photometry are briefly described in \S2. The catalog of variable stars and preliminary statistics of the variable star types are presented in \S3. Our results are summarized in \S4.
 
\section{Observations and data reduction}

\subsection{Observations} 
AST3 \citep{Cui08,Yuan14,Yuan15} was conceived as three telescopes, each equipped with one of three SDSS $g$, $r$ and $i$ filters. Each telescope has an entrance pupil aperture of 0.5\,m and a wide FOV of 4.3\,deg$^2$, and is equipped with a 10K $\times$ 10K frame transfer STA1600FT CCD (Charge Coupled Device) camera. The CCD detector is divided into frame store regions at the top and bottom quarters and an image area in the central half in order to operate in frame transfer mode without a shutter---this is part of our risk-mitigation strategy of eliminating mechanisms as far as possible, since the telescope has to operate entirely remotely for 11 months of the year with no possibility of repairs being carried out. The image area of the CCD has 16 readouts, each with 1320\,$\times$\,2640 pixels, including an overscan region of 180 columns on the readout electronics end. More details about the AST3 CCD performance, data system and survey strategy can be found in \citet{Ma12,Shang12}; Z. Shang et al. (2016, in preparation); Q. Liu et al. (2016, in preparation). For the total of 3523 images obtained of our field, 65\% had exposure times of 30 seconds, and the remainder were 60 seconds. The field is not crowded as the median distance between every star and its nearest neighbor from our reference frame is 11.14 pixels (note that the AST3-1 pixel scale is 1.0 arcseconds/pixel). The stellar brightness profiles had a median FWHM (full width half maximum) of 3.73 pixels. The low level of crowding led us to use aperture photometry rather than PSF-fitting.

\begin{deluxetable}{lrr}
\tablewidth{0pt}
\tablenum{1}
\tablecaption{Log of observations\label{tb:log}}
\tablehead{\colhead{Date} & \colhead{\# images} &
\colhead{Total exp.}\\ \colhead{2012} & & \colhead{time (hr)}}
\startdata
3-28       & 156 &  2.56 \\
4-24       &   6 &  0.02 \\
4-25       & 515 &  6.91 \\
4-26       & 516 &  6.03 \\
4-27       &  58 &  0.83 \\
4-28       & 368 & 5.64 \\
4-29       & 666 & 6.55 \\
4-30       & 750 & 6.25 \\
5-01       & 488 & 4.07\\
{\bf Total} & 3523 & 38.86
\enddata
\end{deluxetable}

The field probes the Galactic disk center at $l=289.6347^{\circ}$, $b= -1.5718^{\circ}$, which was also monitored by the Optical Gravitational Lensing Experiment \citep[OGLE-III; Fig. 1 of ][]{Pietrukowicz13}. Limited data bandwidth \citep[128kbps for our Iridium OpenPort system; ][]{Xu12} from Dome A meant that the raw images were carried back from Dome A on hard disk drives by the 29th Chinese Antarctic Research Expedition (CHINARE) team. The satellite bandwidth is sufficient for transferring only small sections of images and highly-reduced data \citep{Shang12}.  

\subsection{Data reduction}
The preliminary reduction of the raw science images involved crosstalk correction, bias subtraction, dark current subtraction, and flat fielding. The inter-channel interference crosstalk due to the  multi-channel CCD readout, was corrected first. Overscan regions of the 16 readouts were used to correct the corresponding bias. Problems with the CCD's thermoelectric cooler during 2012 meant that the images were subject to high dark current levels, comparable to the sky background. A new method was applied to calculate a dark frame from image pairs. More specifically, our dark frame was derived by combining 230 image pairs (each pair having the same temperature and exposure time), and was scaled to the same temperature and exposure time as the scientific images for dark current correction \citep{Ma14a}.

The flat-fielding of AST3-1's wide field was achieved in two steps. Due to the relatively large 4.3\,deg$^2$ FOV, a sky brightness gradient of 1\% $\sim$~10\% from individual twilight flat-field image remained after pre-processing for crosstalk, overscan, and dark current. The 200 twilight flat-field images were selected to correct the sky brightness gradient. More specifically, for each of the 200 twilight flat-field images, the brightness gradient was first fitted with an empirical function based on the sun's altitude, and the angle between the image and the sun. The gradient was removed by dividing each image by the empirical fit, and the resulting 200 twilight images were then median-combined to obtain a master flat field, which was used for flat-fielding corrections for the science images \citep{Wei14}.

After finishing the preliminary reduction, SExtractor \citep{Bertin96} was applied to perform aperture photometry on all the scientific images. The aperture selection with 4 pixels (SExtractor's MAG\_APER parameter) was adopted as it gave the minimum {\it r.m.s} photometric uncertainties on a test image (B. Ma et al. 2016, in preparation). The photometric uncertainty reached 2~mmag for bright stars $<$13 mag on a typical image.
In order to obtain accurate astrometry, the photometric results were fed into SCAMP \citep{Bertin06} to register the positions of all the images with the PPMX system \cite[Position and Proper Motions eXtended,][]{roser08}. Our photometric calibration was divided into two steps. First, a magnitude difference between an individual image and the reference image (the highest quality image) was obtained by matching $\sim$\,1000 bright, isolated stars. Then, stars in the reference image that were also in the APASS (the AAVSO Photometric All-Sky Survey \footnote{https://www.aavso.org/apass}) catalog \citep{Henden16} were used to adjust the zero point of our magnitudes. APASS is an all-sky photometric survey, conducted in five filters: Johnson B and V, plus Sloan $g$, $r$, $i$ bands. The AAVSO (American Association of Variable Star Observer \footnote{https://www.aavso.org/}) was created by amateur astronomers, and collects and archives variable star observations.

\subsection{Time-series photometry}

During the 8 days of observations, 3,523 images were collected in total for our field and 96,734 bright stars with $S/N > 30$ were found on our reference image, which had the largest number of detections. We then rejected stars that were detected in fewer than 20\% of the images. We are ignoring such objects but they could be interesting. This left a final sample of 92,583 stars.

\begin{figure}[t]
\begin{center}
\includegraphics[width=0.45\textwidth]{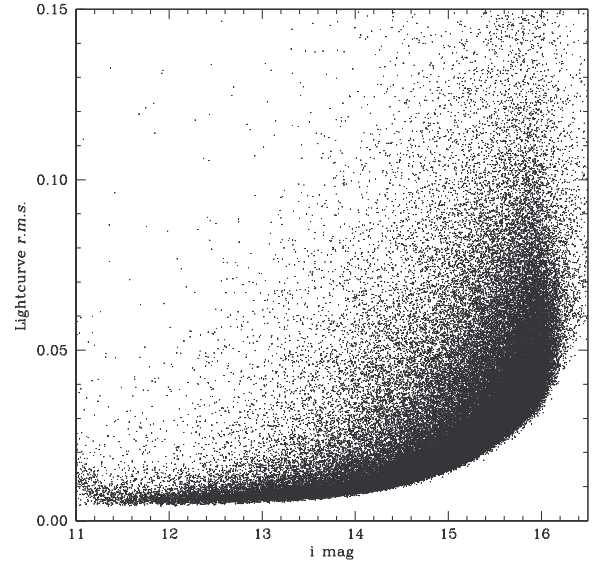}
\caption{{\it r.m.s} magnitude range of light curves of 92583 stars with at least 20\% measurements. \label{fig:magerr}}
\end{center}
\end{figure}

Fig.~\ref{fig:magerr} displays the {\it r.m.s} variations in the light curves of our sample stars as a function of error-weighted magnitude. The magnitude were weighted by their photometric errors after rejecting 3$\sigma$ outliers iteratively until the maximum iteration reaches 10. The {\it r.m.s} is less than 0.02 mag for stars brighter than 15.4 mag. And the {\it r.m.s} is less than 0.05 mag when the magnitudes are smaller than 16.4 mag, which accounts for 78\% stars in our sample.

The photometric errors estimated by SExtractor were under- or overestimated  due to various reasons such as underestimated flat-fielding errors, and less than perfect photometry. We worked around this problem by assuming that the majority of stars are constant and assuming the errors for this majority are roughly Gaussian as the $\chi2/N_{DOF} = 1$ for the constant stars. Following the reference in \citet{Kaluzny98}, we firstly calculated the $\chi2/N_{DOF}$ value for all the stars and then derived a scale factor curve for constant stars. Then we re-scaled all stars by multiplying by this curve. The re-scaled photometric errors were then used in the calculation of the Welch-Stetson variability index $L$ \citep{Welch93,Stetson1996} to find variable candidates for the next section.

\section{Variable Star Catalog and Statistics} 

\subsection{Searching for Variability}

The search for variable stars in our sample was conducted in three steps. Our candidates were initially selected as those stars with statistically significant magnitude variations. Such stars didn't necessarily show a periodic behavior. Then for the selected candidates we conducted a search for periodic behavior. Finally we used visual inspection of the phase-folded light curve and time-series diagram of each candidate to distinguish periodic and aperiodic variables.    

\begin{figure}[t]
\begin{center}
\includegraphics[width=0.45\textwidth]{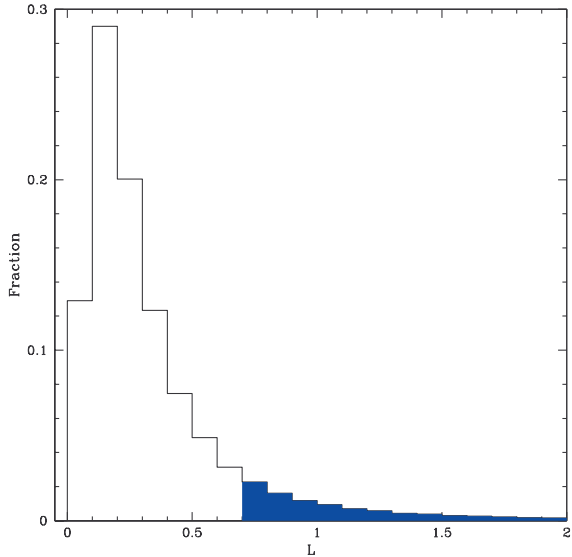}
\caption{Distribution of the Welch-Stetson variability statistic $L$\ \citep{Welch93,Stetson1996} for the 92,583 brightest stars in the AST3-1 sample. \label{fig:lhist}}
\end{center}
\end{figure}

Variable stars exhibit magnitudes variations that can be measured by the Welch-Stetson index \citep{Welch93}, later slightly modified as the $L$ variability index by \citet{Stetson1996}. We calculated $L$ for each light curve in our sample using VARTOOLS \footnote{http://www.astro.princeton.edu/$\sim$jhartman/vartools.html} \citep{vartools,Hartman16}. The resulting $L$ distribution in our sample is shown in Fig.~\ref{fig:lhist}; the overall distribution can be interpreted in terms of two components, one of which -- the one presumably corresponding to non-variable stars -- resembles a Gaussian, while the other -- presumably corresponding to the variables -- behaves rather like an exponential tail. We measured a median value of $L =0.22\pm0.17$ for all stars with $L <0.8$. We used $L\ge 0.65$ (equivalent to a $+2.5\sigma$ selection) as the cut-off for our variable candidates.

We use Lomb-Scargle \citep[][hereinafter LS]{Lomb1976,Scargle1982} and box fitting algorithms \citep[][hereinafter BLS]{Kovacs02} to detect the periods of our variable candidates. The LS method applies the statistical properties of least-squares frequency analysis of unequally spaced data on a series of test periods. We hunted for periods between 0.01 and 10\,d and applied a bin size of 0.01\,d. Periods with $S/N \ge 12 $ in the periodogram were taken to be significant. The BLS method searches for signals characterized by a periodic alternation between two discrete levels with much less time spent at the low-level (occultation) phase. Similar methods (LS \& BLS) for hunting for variable stars, were applied in \citet{Wang11,Wang13a,Yao15}. Surveys such as OGLE II use the Detached Eclipsing Binary Light curve fitter \citep[][hereinafter DEBiL]{Devor05} for finding and analyzing eclipsing binaries in large datasets. We also used the DEBiL code to find the corresponding periods of our variable candidates as an independent check of the periods found by the LS and BLS methods. We visually inspected the phased light curve folded by the periods found from each of the LS, BLS and DEBiL methods, and selected the period that had the smallest $r.m.s.$. Due to the very short observing window (only eight days of observations over 34 days) and observing gaps (when AST3-1 was pointing elsewhere, and avoiding twilight), we found that in some cases periods produced by the LS, BLS or DEBil methods did not produce well-folded phase curves. In these cases we manually adjusted the period to produce the best result.

\par

In order to calculate the uncertainty of the above derived periods, we ran Markov Chain Monte Carlo \citep[MCMC; ][]{mcmc} simulations of the high-order harmonic function as given in equation (\ref{eqn:mcmc}),  
\begin{equation}
F(t)=a_{0}+\sum_{i=1}^{12} \left[ a_{i} \sin (\frac{2 \pi t i}{P})+b_{i}\cos (\frac {2 \pi t i}{P}) \right]
\label{eqn:mcmc}
\end{equation}
based on the detected period using the {\it -nonlinfit} tool of VARTOOLS \citep{Hartman16}. We used a 12th order fit to the light curve, which was able to fit eclipsing binaries with relative deep depths. The number of accepted links in a given fit was set to 1000. The initial guess and step size of each $a_{i}$ and $b_{i}$ were found by fitting the equation (\ref{eqn:mcmc}) to each light curve with a fixed period. The median value of each period and its uncertainty were calculated in the MCMC simulation by again fitting the equation (\ref{eqn:mcmc}) to each light curve, but now with the constraint $0.01 < P < 8.0$.

\begin{figure}[t]
\begin{center}
\includegraphics[width=0.45\textwidth]{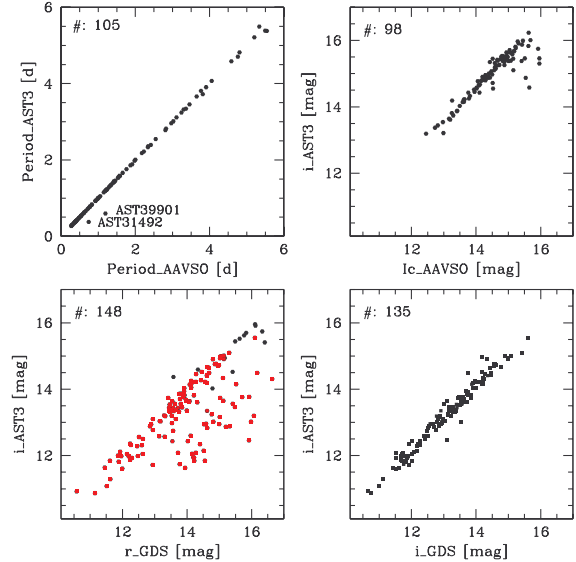}
\caption{Top left: the measured periods for 105 variable stars from AST3-1 compared to the periods given in the AAVSO database; top right: the measured $i$ magnitude for 98 variable stars from AST3-1 compared to the Cousins' $Ic$ magnitude given in the AAVSO database; bottom left: black solid circles show $i$ magnitude for 148 variable stars from AST3-1 compared to the Sloan $r$ magnitude given in the GDS catalog (red squares for 135 out 148 stars with both $r$, $i$ band magnitudes); bottom right: $i$ magnitude for 135 variable stars from AST3-1 compared to the Sloan $i$ magnitude given in the GDS catalog.
  \label{fig:period}}
\end{center}
\end{figure}

Next, we visually inspected the light curve of each variable candidate to search for objects with statistically significant variations in magnitude that did not necessarily show a periodic behavior during our 8-day observations. Our final catalog of variables contains 560 stars in the magnitude range from 10.87 mag to 16.23 mag, 339 of which are new discoveries by AST3-1. Table 2 lists the properties of all the detected variables. Column 1 lists the 2012 AST3 ID; columns 2 and 3 give the right ascension and declination matched to the PPMX system; column 4 contains the weighted mean $i$-band magnitude; column 5 gives the $L$ value;  column 6 lists the most significant median period (top left panel in Fig.~\ref{fig:period}) in the MCMC simulation, and its standard deviation in brackets in unit of $10^{-9}$~d (when applicable); column 7 specifies the minimum $\chi^2$ per degree of freedom in the MCMC simulation; column 8 gives the peak-to-peak amplitude of variation; column 9 gives the time of the first minimum light contained in our observations (only for the periodic variables); column 10 contains a tentative classification of the variable \footnote{http://www.sai.msu.su/gcvs/gcvs/iii/vartype.txt}, where possible; column 11 has additional information, including previous identification of the variable from the all-sky automated survey run by AAVSO \citep{Pojmanski2005,Watson16} or inclusion in the variable stars of OGLE-III \citep{Samus2009} or the Bochum Survey of the Southern Galactic Disk \citep[hereinafter GDS]{Hackstein15}. 

 We matched our final catalog of 560 stars with the AAVSO database and GDS catalog, with a matching radius of 15 arcseconds. This initially resulted in 231 variables stars being previously known. There were 116 variable stars in common between our survey and the AAVSO database, and 104 of these had magnitudes in the Cousins' infra-red $Ic$ band \citep{Watson16}. There are 152 variable stars in common between our catalog and the GDS database with magnitude values in Sloan $r$ band; 139 of these also had magnitude measurements in Sloan $i$ band \citep{Hackstein15}, which is similar to our filter. There are 37 variables that appeared in both the AAVSO database and the GDS catalog. We double-checked our initial matching results by measuring the magnitude difference $\Delta$\,mag our observations and $Ic$ (AAVSO) and $i$ (GDS). For the 104 stars in common with the AAVSO database, we measured a median value of $\Delta~mag=(i-Ic)=0.553~\pm0.316$ and applied 3$\sigma$ outliers rejection to the $\Delta~mag$ and also required that the $Ic$ magnitude should be less than 16.5~mag to exclude 6 stars (AST11766, 12622, 37735, 64879, 74215 and 89418) from the initial common sample. The $i$ vs. $Ic$ magnitude diagram for the remaining 98 variable stars is shown in the right panel of Fig.~\ref{fig:period}. We excluded 4 outliers (AST05375, 57964, 71042 and 83522) with the same method applied to the 139 stars common between our catalog and GDS. $i$ vs. $r$ for 148 variables we have in common with GDS is shown in the bottom left panel of Fig.~\ref{fig:period}; $i$ vs. $i$ for 135 variables is shown in the bottom right panel.
   
\begin{figure}[htb] 
\begin{center}
\includegraphics[width=0.45\textwidth]{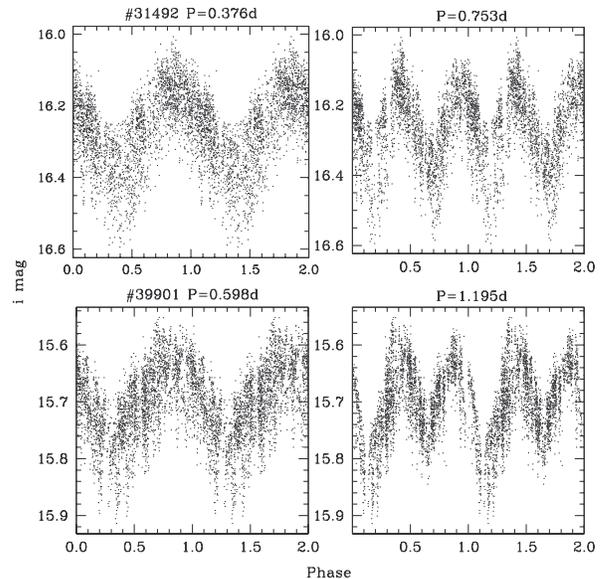}
\caption{Phased light curves were folded by periods from AST3-1 (left panels) and from the AAVSO database (twice that from AST3-1, right panels) for AST31492 and 39901; we believe that the AST3-1 periods are more likely to be correct.
   \label{fig:p1p2}}
\end{center}
\end{figure}

  Our final catalog shows that we have rediscovered 221 previously known variable stars; 96 of these are found in both the AAVSO database and the OGLE-III survey \citep{Pietrukowicz13}; an additional 14 stars were in the AAVSO database but not in the OGLE-III survey; the remaining 111 stars were found in the GDS catalog. Due to our short observing window, we could not determine the periods of three variables: AST13431, 49035, and 83717. In addition, AST43064 and 62877 are listed in the AAVSO database without a period, and all known variables from the GDS catalog had no measurements of their periods until now. The estimated periods for the remaining 105 variables from AST3-1 are highly consistent with those given in the AAVSO database except for five variables: AST31492, 38531, 39901, 59773, and 68860, which have AST3-1 periods that are half of the AAVSO values. Our period detection methods (LS, BLS) are sensitive to detect sine wave or box-like signals. Our periods should be double for binary stars to show the primary and secondary eclipsing light variabilities, i.e., binaries AST38531, 59773 and 68860. While for another two periodic variable stars AST31492 and 39901, we believe our periods are correct (Fig.~\ref{fig:p1p2}). The final period-period diagram from AST3-1 and AAVSO database is shown in top left panel of Fig.~\ref{fig:period}.

\begin{figure}[htb] 
\begin{center}
\includegraphics[angle=-90,width=0.45\textwidth]{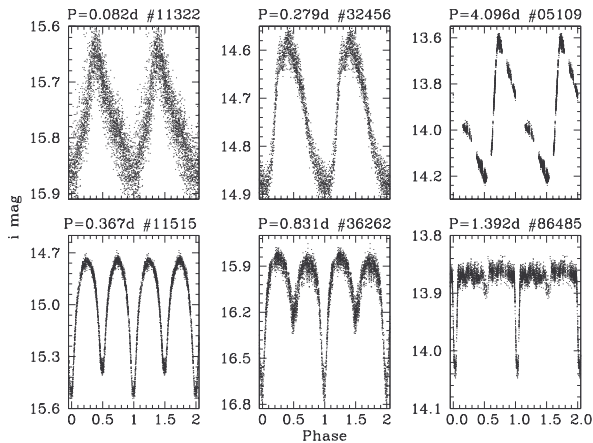}
\caption{Phased light curves for six typical periodic variable stars. The periods and AST IDs are listed above each panel. Top row (from left to right): $\delta$ Scuti, RRLyr c-type star, and $\delta$ Cepheid; bottom row (from left to right): eclipsing binaries of W UMa-type (EW), $\beta$ Lyrae-type (EB), and Algol-type (EA) configurations. 
  \label{fig:phase}}
\end{center}
\end{figure}

We classified half of the variables into binaries, 17\% of them into $\delta$~Scuti, $\gamma$~Doradus,  $\delta$ Cephei, RR Lyrae, and unclassified periodic or multiple periodic variables. Due to our short observing window, we could not detect the periods of the one third of the variables that passed our selection as a result of their significant time series variability. Table~3 contains approximate statistics for the different types, if possible. The folded light curves of the representative periodic variables are showed in Figs.~\ref{fig:phase} -~\ref{fig:mp}, while the representative light curves of aperiodic variables are showed in Fig.~\ref{fig:var}. The time-series data of all 560 variable stars will be available in machine readable format through the VizieR Online Data Catalog \footnote{\url{http://vizier.u-strasbg.fr/}}.

\subsection{Types of variables found by AST3-1}

\subsubsection{Eclipsing binaries}

 Eclipsing binaries can be classified into three broad categories based on the shape of their light curves: Algol-type eclipsing systems (EAs), $\beta$ Lyrae-type eclipsing systems (EBs) and W Ursae Majoris-type eclipsing variables (EWs). The EA systems have obviously different depths between the primary and secondary minima, and have clearly defined times for the beginning and ending of the eclipses; EA systems are often but not always a detached eclipsing system \citep{Catelan15}, although the prototype of the class, Algol, is believed to be a semi-detached system \citep{Soderhjelm80,Kolbas15}. The EB systems show a continuous change in brightness and have a deeper primary depth than that of the secondary. The EW systems also show a continuous change in brightness and have an almost equal or non-obvious varying depth between the primary and secondary minima. The EW systems consist of two components almost in contact and thus have periods generally shorter than 1 day. Among our sample, there are 127 EWs, 33 EBs, 138 EAs and 65 stars in question are probably distributed into several different variability classes, including eclipsers, pulsators, and others, which are separated by a pipe symbol ``$\mid$'' in Table 2. In total, we have detected 285 binaries and 143 ones are new detections from our data. Of the 339 new variables, 42\% belong to the classes of eclipsing binary stars.

\begin{figure}[t]
\begin{center}
\includegraphics[width=0.45\textwidth]{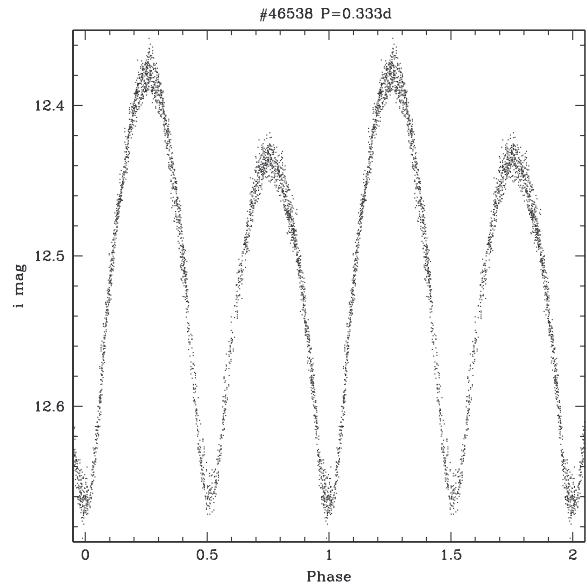}
\caption{The phased diagram of AST46538, which has the most significant O'Connell effect in our sample. Its ID and orbital period are marked in the title. \label{fig:ECOC}}
\end{center}
\end{figure}

There are 34 interesting EW or EB binaries among our 285 detections which show O'Connell effects, i.e., the two successive out-of-eclipse maxima have unequal height in the light curves \citep{OConnell1951,Milone1968,Nataf2010}. The O'Connell effect can be explained by the interaction of circumstellar material with the binary components \citep{Liu2003}. The interaction model suggests that the O'Connell effect is most obvious in late type and/or short period binaries. In our sample, a short-period EW binary AST46538 exhibits the most obvious O'Connell effect (Fig.~\ref{fig:ECOC}), and has a magnitude difference of 0.06 mag in $i$ band between the first and second maximum out-of-eclipse brightness. AST46538 appears in the UCAC4 catalog \citep[Fourth U.S. Naval Observatory CCD Astrograph Catalog;][]{Zacharias2013} and has $B=13.697$~mag and $V=12.916$~mag. It also appears in the 2MASS All-Sky catalog of Point Sources \citep{Cutri03} and has $J, H, K$ of $11.311\pm0.026, 10.941\pm0.029, 10.842\pm0.027$~mag., respectively. In order to estimate the color excess $E(B-V)$ of AST46538, we compared its color-color diagrams ($V-J$ vs. $B-V$, $V-H$ vs. $B-V$, $V-K$ vs. $B-V$, $J-H$ vs. $B-V$ and $H-K$ vs. $B-V$) with the intrinsic color-color diagrams for main sequence stars \footnote{http://www.stsci.edu/$\sim$inr/intrins.html} \citep{Fitzgerald1970,Ducati01} and we found that its interstellar extinction can be neglected. Thus the color term of AST46538 from UCAC4 and 2MASS catalogs can be taken as its intrinsic color to estimate its spectral type, which is equivalent to spectral type G8 \citep{Fitzgerald1970,Ducati01}. We find that the spectral type G8 and orbital period 0.333d of AST46538 are close to that of the W UMa binary YY Eri with spectral type G5V and orbital period 0.322d \citep{Liu2003}; thus, both should exhibit similar O'Connell effects based on the model of \citet{Liu2003}. Indeed, we find consistent O'Connell effects in AST46538 and YY Eri: for YY Eri, the modeled bolometric magnitude difference is 0.07 mag and the observed magnitude difference in $V$ band is 0.04 mag \citep{Liu2003}, while for AST46538, the observed magnitude difference in $i$ band is 0.06 mag. More observations are needed to double check the above analysis for AST46538.

\ \par

\subsubsection{Pulsating variable stars}
Pulsating variable stars exhibit periodic expansion and contraction (radially or non-radially) of their surface layers \citep{Catelan15}. The pulsating variable stars are classified into many types based on their period, amplitude, light curve shape, evolutionary status and so on. A more complete classification for all kinds of pulsating variable stars ($\delta$~Scuti, $\gamma$~Doradus, RR Lyrae stars, Cepheids and so on) can be found at \citet{Catelan15}. $\delta$ Scuti variables are late A- and early F-type stars situated in the instability strip on or above the main sequence in the HR Diagram. Their typical pulsation periods are found to be in the range of 0.02d to 0.25\,d \citep{Breger2000}. $\gamma$~Doradus stars locate in the similar position in the instability strip as the $\delta$~Scuti stars but with relatively larger pulsating periods ranging from 0.3\,d to 3\,d \citep{Cuypers2009}. RR Lyrae stars are radially pulsating giant stars with spectral type A to F with periods from $\sim$ 0.2 to $\sim$1\,d \citep{Smith04}. Most RR Lyrae stars are pulsating in the radial fundamental mode (RRab stars) and the first overtone mode (RRc stars).

\begin{figure}[htb] 
\begin{center}
\includegraphics[width=0.45\textwidth]{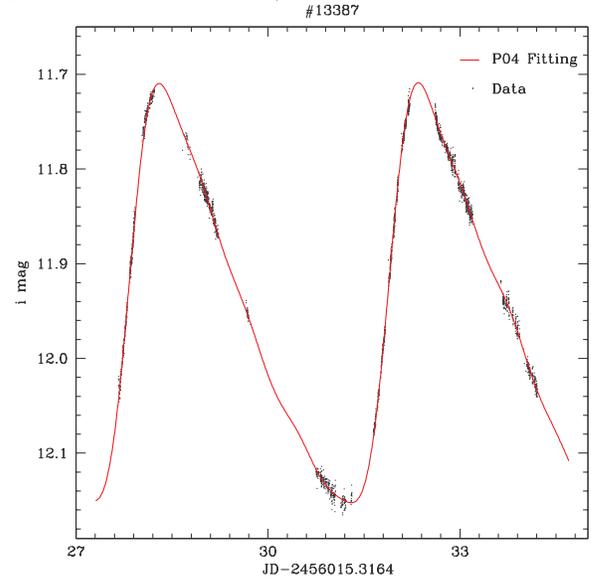}
\caption{Observed light curves of the known $\delta$ Cepheid GS Car (AST13387) in $i$ band (black points) and the fitting curves with the fundamental frequency and the harmonics (red curves).   
  \label{fig:dcep}}
\end{center}
\end{figure}

 Cepheid variables obey the period-luminosity relation and are divided into two subclasses---type I and type II Cepheids \citep{Catelan15}---based on their masses, ages and evolutionary states. The period-luminosity relation shows that there is a subtype between I and II, called Anomalous Cepheids \citep[Figure 7.3 of ][]{Catelan15}. The majority of $\delta$~Cepheid variables show a large light variation and a rapid rise to maximum and a slow decline back to minimum (i.e., Fig.~\ref{fig:dcep}), which is similar to a RR Lyrae star \citep{Schmidt04,Soszyski08a}. There are $\delta$ Cepheids with lower amplitudes ($<0.5$~mag. in $V$) but they have symmetrical light curves and shorter periods \citep[$<7$ days; ][]{Catelan15}. Type II Cepheids generally shows a relatively broad maximum and a symmetric minimum \citep{Schmidt04} and they have periods of $\sim$~0.8\,-\,35 days and light amplitudes from 0.3 to 1.2 mag in $V$ band \footnote{http://www.sai.msu.su/gcvs/gcvs/iii/vartype.txt}. Anomalous Cepheids show the similar light-curve morphology as RR Lyrae variables with periods shorter than two days and the majority of them show a small bump before the rise to maximum \citep{Soszyski08b}. Based on the light-curve morphology, 12 variable candidates were classified as $\delta$ Cepheids as they have higher light amplitudes ($>$ 0.3 mag.), larger periods ($\gtrsim$ 2~days) and also exhibit a rapid rise to maximum and a slow decline to minimum. Another 13 variables (AST04480, 09282, 35419, 35518, 40551, 42471, 49241, 53255, 67933, 75631, 79599, 81000, and 84533) also show the morphology of a fast rise and a slow fall, and we have classified them as type ``PER''  (\S3.2.3) since their $i$ amplitudes are less than 0.3 mag.\ and they have no bumps before the rise to maximum.

In our sample, there are 27 pulsators, which are classified into 10 $\delta$~Scuti stars, 2 $\gamma$~Doradus stars, 12 $\delta$~Cepheids, and 3 RR Lyraes. For example, AST13387 is the previously known $\delta$~Cepheid GS Car. Its phased light curve in $i$ band can be well fitted by the Fourier decompositions with a fundamental frequency of 0.245616\,d$^{-1}$ and the harmonics of 0.497130, 0.736473, 0.984846, 1.225154\,d$^{-1}$) in order of decreasing amplitude values (the red curves in Fig.~\ref{fig:dcep}).    


\subsubsection{Other types of variable stars}

\begin{figure}[t] 
\begin{center}
\includegraphics[width=0.45\textwidth]{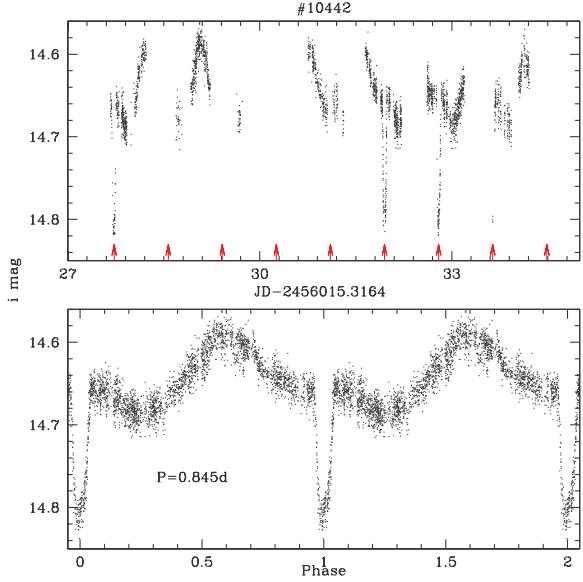}
\caption{Top panel: Light curve of AST10442 in the $i$ band, showing an RS CVn-like pattern with a period of 0.845\,d, and a prominent primary minimum clearly detected (marked with arrows in the top plot). Bottom panel: Phased light curve of AST10442 folded by the period of 0.845\,d. 
  \label{fig:mp}}
\end{center}
\end{figure}

For the unclassified 248 variable stars we detected the main periods for 67 of them (periodic and multi-periodic). If the phased light curve folded by its main period was significantly scattered, we classify it as a multi-periodic type variable star, otherwise we classify it as a periodic one. For the remaining 181 unclassified objects we could not detect a period due to the short observing window (8 days spanning 34 days). In our sample, we detected a new complex binary system, AST10442 (Fig.~\ref{fig:mp}). The system has a period of 0.845\,d, and an $i$ magnitude of 14.64. The system, which presents RS CVn-like light curve morphology (obvious variability when out of eclipse), shows a primary depth of 0.15 mag but does not show the secondary eclipse on the folded light curve (bottom panel of Fig.~\ref{fig:mp}). AST10442 appears in the SPM4 catalog \citep{Girard2011} and has $B = 16.59$ mag and $V = 15.24$ mag. It also appears in the 2MASS All-Sky catalog of Point Sources \citep{Cutri03} and has $J, H, K$ of $13.254\pm0.029, 12.728\pm0.022, 12.548\pm0.030$~mag., respectively. We have done the same analysis (color-color diagrams) as the AST46538 and found its interstellar extinction can also be neglected. Based on the empirical formula $T_{\rm eff}=\frac{8540}{(B-V)+0.865}$\citep{Swamy96,Zong15}, color $B-V=1.35$~mag is equivalent to $T_{\rm eff} \approx 3850$~K. Such low $T_{\rm eff}$ suggests that AST10442 has a spectral type of K5 or M0. If it is a giant, its orbital velocity could be $>500$ km/s from Figure 4 of \citet{Gaulme2013}. To our knowledge, no published spectroscopic observations have been performed of this interesting system to date. We have obtained 2 hours on Gemini South to carry out spectroscopic observations of this system, in order to measure its maximum radial velocities at phases 0.25 and 0.75.

\begin{figure}[htb] 
\begin{center}
\includegraphics[width=0.45\textwidth]{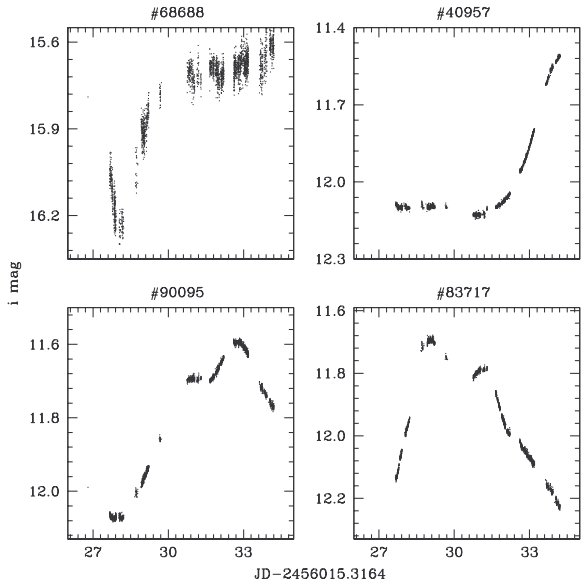}
\caption{Light curves of four unclassified aperiodic variable stars with the largest variability amplitudes.
  \label{fig:var}}
\end{center}
\end{figure}

In addition, out of the 181 aperiodic variable stars, four variable stars AST68688, 40957, 90095 and 83717 with the largest amplitudes in Table~2, are showed in the four panels of Fig.~\ref{fig:var}. AST68688 and 40957 have shown fast rising variability, i.e., $\sim$\,0.6 mag in 3 days (top panels of Fig.~\ref{fig:var}). AST90095 shows a brightness plateau during its ascending stage (bottom left panel of Fig.~\ref{fig:var}); its variability was detected in the GDS catalog, based on sparse data with big gaps \citep{Hackstein15}. The new variable AST83717 shows a secondary maximum during its descending stage after its $i$ band maximum brightness (bottom right panel of Fig.~\ref{fig:var}). More observations are needed for further investigations into these four interesting objects.


\LongTables


\section{Summary}
  We have presented the analysis of $i$ band images survey from the AST3-1 telescope towards one Galactic disk field centered at $l=289.6347^{\circ}$, $b= -1.5718^{\circ}$. 560 variable stars were detected in the field from the time series photometry of 92,583 stars with $i$ magnitude $\leq$ 16.5 mag during the eight days of observations. Multiple methods (LS, BLS, DEBiL and visual inspection) were used to look for the initial periods and we adopted the one that gave the smallest scatter in the phase-folded light curves. We used MCMC simulations based on the harmonic function of the primary period for each variable star to estimate the uncertainty of the period and its median value. For the previously known periodic variable stars, the median periods from the MCMC simulations are highly consistent with those given in the AAVSO database, see Fig.~\ref{fig:period}. 

  We tentatively classified the 560 variables into 285 eclipsing binaries (EWs, EBs, EAs), 27 pulsating variable stars ($\delta$~Scuti, $\gamma$~Doradus, $\delta$~Cepheid variable and RR Lyrae stars) and 248 other types of variables (unclassified periodic, multi-periodic and aperiodic variable stars). Out of the 560 variables, 339 (61\%) are new detections from our data; 42\% of the new detections are eclipsing binary stars. We found 34 eclipsing binaries that show O'Connell effects. The interaction between circumstellar matter and the binary components may offer an explanation for the O'Connell effect (see the discussion on AST46538 in \S3.2.1). We also found one aperiodic variable that shows a plateau light curve and another one that shows a secondary maximum after peak brightness. Among our newly discovered variables, we found one with complex behaviour showing a binary system with RS CVn-like light curve morphology; we are in the process of obtaining spectroscopic follow-up observations of this object using the Gemini South telescope. All the time-series data of the variable stars will be available via the VizieR Online Data Catalog \footnote{\url{http://vizier.u-strasbg.fr/}}.

\acknowledgements

We thank M$\acute{a}$rcio Catelan for helpful discussions and Ricardo Salinas when applying for spectral follow-up observations of AST10442. We appreciate comments from the anonymous referees, Lucas Macri for discussions on the Welch-Stetson variability $L$ and Joel Hartman for the discussions on the period uncertainty. This work was supported by the National Basic Research Program (973 Program) of China (Grant Nos.\ 2013CB834901, 2013CB834900, 2013CB834903), the Chinese Polar Environment Comprehensive Investigation \& Assessment Program (Grand No.\ CHINARE2016-02-03-05), the National Natural Science Foundation of China (NSFC grants 11303041, 11203039, 11273019, 11273025, 11403048, 11473038, 11673003), the Australian Antarctic Division, and the National Collaborative Research Infrastructure Strategy (NCRIS). LZW acknowledges the Chinese Academy of Sciences (CAS), through a grant to the CAS South America Center for Astronomy (CASSACA) in Santiago, Chile and support by the One-Hundred-Talent program of the Chinese Academy of Sciences (034031001). LFW acknowledges the Strategic Priority Research Program "The Emergence of Cosmological Structures" of the Chinese Academy of Sciences, Grant No. XDB09000000. JNF acknowledges the support from the Joint Fund of Astronomy of National Natural Science Foundation of China (NSFC) and Chinese Academy of Sciences through the grant U1231202, the National Basic Research Program of China (973 Program 2014CB845700), and the LAMOST FELLOWSHIP supported by Special Funding for Advanced Users, budgeted and administrated by Center for Astronomical Mega-Science, Chinese Academy of Sciences (CAMS). The authors deeply appreciate the great efforts made by the 24--32th Dome A expedition teams, who provided invaluable assistance to the astronomers that set up and maintained the AST3 and the PLATO-A system.


\bibliography{paper}{}
\bibliographystyle{apj}

\clearpage
 \LongTables
\begin{deluxetable}{lrrr}
\tablewidth{0pt}
\tablenum{4}
\tablecaption{Identifications for the 221 variables in common \label{tb:match}}
\tablehead{\colhead{(1)} & \colhead{(2)} & \colhead{(3)} & \colhead{(4)}  \\
  \colhead{AST3} & \colhead{AAVSO} & \colhead{GDS}& \colhead{OGLE-III or others} }
\startdata
AST01736 & $\dots$ &  GDS\_J1050304-613526 &              $\dots$ \\
AST02497 &  310338 &               $\dots$ & OGLE-GD-ECL-03652    \\
AST03182 &  310357 &               $\dots$ & OGLE-GD-ECL-03671    \\
AST04554 & $\dots$ &  GDS\_J1050526-615847 &              $\dots$ \\
AST05016 & $\dots$ &  GDS\_J1050463-602841 &              $\dots$ \\
AST05512 &  310418 &               $\dots$ & OGLE-GD-ECL-03733    \\
AST06264 &  310438 &  GDS\_J1051045-612147 & OGLE-GD-ECL-03753    \\
AST06935 &  310450 &               $\dots$ & OGLE-GD-ECL-03765    \\
AST07839 & $\dots$ &  GDS\_J1051142-614524 &              $\dots$ \\
AST08666 &   92320 &  GDS\_J1051153-603208 & ASAS J105115-6032.1  \\
AST10372 & $\dots$ &  GDS\_J1051315-600945 &              $\dots$ \\
AST11515 & $\dots$ &  GDS\_J1051453-614613 &              $\dots$ \\
AST11525 &  310601 &  GDS\_J1051435-610220 & OGLE-GD-ECL-03919    \\
AST12715 &  310634 &               $\dots$ & OGLE-GD-ECL-03957    \\
AST13387 &    5954 &  GDS\_J1051541-612802 & GS Car               \\
AST13431 &  226193 &               $\dots$ & [PMF2009] V041       \\
AST13451 &  310655 &               $\dots$ & OGLE-GD-ECL-03980    \\
AST14578 &  310681 &  GDS\_J1052054-612031 & OGLE-GD-ECL-04022    \\
AST15280 & $\dots$ &  GDS\_J1052104-603541 &              $\dots$ \\
AST16017 & $\dots$ &  GDS\_J1052179-612010 &              $\dots$ \\
AST16336 & $\dots$ &  GDS\_J1052096-612546 &              $\dots$ \\
AST16539 &  226339 &  GDS\_J1052229-613208 & [PMF2009] V114       \\
AST16578 & $\dots$ &  GDS\_J1052152-602413 &              $\dots$ \\
AST16670 & $\dots$ &  GDS\_J1052165-602348 &              $\dots$ \\
AST16693 & $\dots$ &  GDS\_J1052219-603335 &              $\dots$ \\
AST17775 &  358801 &               $\dots$ & OGLE-GD-CEP-0005     \\
AST17938 &  310754 &               $\dots$ & OGLE-GD-ECL-04118    \\
AST17970 & $\dots$ &  GDS\_J1052246-603438 &              $\dots$ \\
AST18900 &  310774 &               $\dots$ & OGLE-GD-ECL-04146    \\
AST19397 & $\dots$ &  GDS\_J1052465-615633 &              $\dots$ \\
AST19571 &  310815 &  GDS\_J1052504-615557 & OGLE-GD-ECL-04193    \\
AST20003 & $\dots$ &  GDS\_J1052484-612033 &              $\dots$ \\
AST20657 &  310822 &  GDS\_J1052518-612125 & OGLE-GD-ECL-04200    \\
AST20751 &  310834 &               $\dots$ & OGLE-GD-ECL-04212    \\
AST21008 & $\dots$ &  GDS\_J1052487-603916 &              $\dots$ \\
AST21898 &  310869 &  GDS\_J1053033-613822 & OGLE-GD-ECL-04249    \\
AST22118 & $\dots$ &  GDS\_J1053013-610728 &              $\dots$ \\
AST22962 &  310898 &               $\dots$ & OGLE-GD-ECL-04280    \\
AST23445 & $\dots$ &  GDS\_J1053079-600853 &              $\dots$ \\
AST23515 &  310915 &               $\dots$ & OGLE-GD-ECL-04309    \\
AST23549 & $\dots$ &  GDS\_J1053169-621112 &              $\dots$ \\
AST23732 & $\dots$ &  GDS\_J1053269-625034 &              $\dots$ \\
AST24162 & $\dots$ &  GDS\_J1053208-611234 &              $\dots$ \\
AST24483 &  310927 &  GDS\_J1053209-605115 & OGLE-GD-ECL-04325    \\
AST24595 &  310958 &               $\dots$ & OGLE-GD-ECL-04363    \\
AST24761 & $\dots$ &  GDS\_J1053314-620336 &              $\dots$ \\
AST24936 & $\dots$ &  GDS\_J1053343-621959 &              $\dots$ \\
AST25082 & $\dots$ &  GDS\_J1053201-602952 &              $\dots$ \\
AST25633 & $\dots$ &  GDS\_J1053332-612915 &              $\dots$ \\
AST25740 &  226643 &               $\dots$ & [PMF2009] V266       \\
AST26037 & $\dots$ &  GDS\_J1053336-610336 &              $\dots$ \\
AST26542 & $\dots$ &  GDS\_J1053357-612416 &              $\dots$ \\
AST26602 &    5982 &  GDS\_J1053415-613654 & IL Car               \\
AST27264 & $\dots$ &  GDS\_J1053355-602114 &              $\dots$ \\
AST27396 &  311017 &  GDS\_J1053454-611535 & OGLE-GD-ECL-04451    \\
AST27446 & $\dots$ &  GDS\_J1053405-604122 &              $\dots$ \\
AST27616 &  311038 &  GDS\_J1053518-614300 & OGLE-GD-ECL-04478    \\
AST28591 & $\dots$ &  GDS\_J1053575-614028 &              $\dots$ \\
AST28592 & $\dots$ &  GDS\_J1053555-612813 &              $\dots$ \\
AST28694 & $\dots$ &  GDS\_J1053474-603313 &              $\dots$ \\
AST29379 & $\dots$ &  GDS\_J1053508-601527 &              $\dots$ \\
AST29542 &  311073 &               $\dots$ & OGLE-GD-ECL-04521    \\
AST30846 & $\dots$ &  GDS\_J1054054-603728 &              $\dots$ \\
AST31492 &  311144 &               $\dots$ & OGLE-GD-ECL-04596    \\
AST31705 &  311154 &               $\dots$ & OGLE-GD-ECL-04606    \\
AST33265 &    5857 &  GDS\_J1054268-612049 & CC Car               \\
AST35554 &  311250 &               $\dots$ & OGLE-GD-ECL-04704    \\
AST35618 & $\dots$ &  GDS\_J1054426-604647 &              $\dots$ \\
AST36262 & $\dots$ &  GDS\_J1054491-604040 &              $\dots$ \\
AST37489 &  311315 &               $\dots$ & OGLE-GD-ECL-04769    \\
AST37704 & $\dots$ &  GDS\_J1055143-623605 &              $\dots$ \\
AST37940 &  311326 &  GDS\_J1055106-613508 & OGLE-GD-ECL-04780    \\
AST38214 & $\dots$ &  GDS\_J1055183-622619 &              $\dots$ \\
AST38428 & $\dots$ &  GDS\_J1055119-614224 &              $\dots$ \\
AST38531 &  311348 &               $\dots$ & OGLE-GD-ECL-04803    \\
AST38989 &  311335 &               $\dots$ & OGLE-GD-ECL-04789    \\
AST39051 & $\dots$ &  GDS\_J1055164-613623 &              $\dots$ \\
AST39579 &  311340 &               $\dots$ & OGLE-GD-ECL-04794    \\
AST39766 & $\dots$ &  GDS\_J1055076-600437 &              $\dots$ \\
AST39901 &  311400 &               $\dots$ & OGLE-GD-ECL-04858    \\
AST39979 &    5983 &  GDS\_J1055097-603251 & IM Car               \\
AST40137 & $\dots$ &  GDS\_J1055312-621143 &              $\dots$ \\
AST40536 &  311382 &  GDS\_J1055215-605455 & OGLE-GD-ECL-04840    \\
AST40655 &  311441 &               $\dots$ & OGLE-GD-ECL-04899    \\
AST40671 &  311383 &  GDS\_J1055216-605032 & OGLE-GD-ECL-04841    \\
AST40766 &  311387 &  GDS\_J1055231-605051 & OGLE-GD-ECL-04845    \\
AST40904 &  311392 &               $\dots$ & OGLE-GD-ECL-04850    \\
AST40957 & $\dots$ &  GDS\_J1055122-600930 &              $\dots$ \\
AST41209 & $\dots$ &  GDS\_J1055237-603407 &              $\dots$ \\
AST41484 & $\dots$ &  GDS\_J1055396-621808 &              $\dots$ \\
AST41594 &  311497 &               $\dots$ & OGLE-GD-ECL-04955    \\
AST41650 & $\dots$ &  GDS\_J1055417-622358 &              $\dots$ \\
AST41745 &  311391 &               $\dots$ & OGLE-GD-ECL-04849    \\
AST42287 &  311470 &               $\dots$ & OGLE-GD-ECL-04928    \\
AST42633 &  311469 &               $\dots$ & OGLE-GD-ECL-04927    \\
AST42854 & $\dots$ &  GDS\_J1055451-614004 &              $\dots$ \\
AST42869 &  311535 &               $\dots$ & OGLE-GD-ECL-04993    \\
AST43064 &   56994 &               $\dots$ & NSV 18559            \\
AST43350 &  311558 &               $\dots$ & OGLE-GD-ECL-05017    \\
AST43409 & $\dots$ &  GDS\_J1055342-600459 &              $\dots$ \\
AST43421 &  311472 &               $\dots$ & OGLE-GD-ECL-04930    \\
AST43594 &  311494 &               $\dots$ & OGLE-GD-ECL-04952    \\
AST43846 & $\dots$ &  GDS\_J1056058-622946 &              $\dots$ \\
AST44142 &  311616 &  GDS\_J1056002-615337 & OGLE-GD-ECL-05075    \\
AST45236 &  311601 &               $\dots$ & OGLE-GD-ECL-05060    \\
AST45483 &  311620 &               $\dots$ & OGLE-GD-ECL-05079    \\
AST45682 & $\dots$ &  GDS\_J1055547-603155 &              $\dots$ \\
AST46247 &  311651 &               $\dots$ & OGLE-GD-ECL-05110    \\
AST46538 & $\dots$ &  GDS\_J1056047-604149 &              $\dots$ \\
AST46624 & $\dots$ &  GDS\_J1056072-605408 &              $\dots$ \\
AST46738 & $\dots$ &  GDS\_J1056223-615935 &              $\dots$ \\
AST46971 & $\dots$ &  GDS\_J1056028-601408 &              $\dots$ \\
AST48664 & $\dots$ &  GDS\_J1056375-621855 &              $\dots$ \\
AST49035 &  311772 &  GDS\_J1056276-605613 & OGLE-GD-ECL-05232    \\
AST50267 &  311762 &               $\dots$ & OGLE-GD-ECL-05222    \\
AST50549 &  311789 &               $\dots$ & OGLE-GD-ECL-05249    \\
AST50581 & $\dots$ &  GDS\_J1056268-601705 &              $\dots$ \\
AST51624 &  311830 &  GDS\_J1056353-601239 & OGLE-GD-ECL-05291    \\
AST53357 &  311884 &  GDS\_J1056468-601442 & OGLE-GD-ECL-05346    \\
AST53821 & $\dots$ &  GDS\_J1056563-604013 &              $\dots$ \\
AST53841 & $\dots$ &  GDS\_J1057193-623419 &              $\dots$ \\
AST54338 &  311973 &               $\dots$ & OGLE-GD-ECL-05436    \\
AST54349 &  311929 &               $\dots$ & OGLE-GD-ECL-05392    \\
AST54537 &  311991 &  GDS\_J1057065-610112 & OGLE-GD-ECL-05454    \\
AST54590 &  312031 &  GDS\_J1057151-613403 & OGLE-GD-ECL-05494    \\
AST55024 &  311997 &               $\dots$ & OGLE-GD-ECL-05460    \\
AST55167 & $\dots$ &  GDS\_J1057307-623116 &              $\dots$ \\
AST55334 & $\dots$ &  GDS\_J1057013-601502 &              $\dots$ \\
AST56058 &  312096 &  GDS\_J1057298-615200 & OGLE-GD-ECL-05560    \\
AST56624 & $\dots$ &  GDS\_J1057448-623935 &              $\dots$ \\
AST56993 & $\dots$ &  GDS\_J1057296-613940 &              $\dots$ \\
AST57491 &  312051 &               $\dots$ & OGLE-GD-ECL-05514    \\
AST59186 &  312200 &               $\dots$ & OGLE-GD-ECL-05667    \\
AST59773 &  312166 &               $\dots$ & OGLE-GD-ECL-05632    \\
AST60119 & $\dots$ &  GDS\_J1057348-602105 &              $\dots$ \\
AST61176 &  312195 &               $\dots$ & OGLE-GD-ECL-05662    \\
AST61299 & $\dots$ &  GDS\_J1058171-623624 &              $\dots$ \\
AST61514 & $\dots$ &  GDS\_J1057588-614517 &              $\dots$ \\
AST61607 &  312291 &               $\dots$ & OGLE-GD-ECL-05761    \\
AST61969 &  312222 &  GDS\_J1057573-605700 & OGLE-GD-ECL-05690    \\
AST62053 &    5772 &  GDS\_J1058105-615457 & SS Car               \\
AST62121 &  312262 &  GDS\_J1058047-612228 & OGLE-GD-ECL-05732    \\
AST62331 &  312278 &               $\dots$ & OGLE-GD-ECL-05748    \\
AST62372 &  312176 &  GDS\_J1057488-601234 & OGLE-GD-ECL-05642    \\
AST62838 & $\dots$ &  GDS\_J1058028-610857 &              $\dots$ \\
AST62877 &  358849 &               $\dots$ & OGLE-GD-RRLYR-0020   \\
AST63063 &  312274 &               $\dots$ & OGLE-GD-ECL-05744    \\
AST63331 &  312220 &               $\dots$ & OGLE-GD-ECL-05688    \\
AST64031 &  312282 &               $\dots$ & OGLE-GD-ECL-05752    \\
AST64308 & $\dots$ &  GDS\_J1058215-613132 &              $\dots$ \\
AST64778 & $\dots$ &  GDS\_J1058176-611203 &              $\dots$ \\
AST65679 &  312442 &               $\dots$ & OGLE-GD-ECL-05913    \\
AST66017 & $\dots$ &  GDS\_J1058445-620806 &              $\dots$ \\
AST66226 &  312330 &               $\dots$ & OGLE-GD-ECL-05800    \\
AST67325 &  312421 &  GDS\_J1058341-610153 & OGLE-GD-ECL-05892    \\
AST67686 & $\dots$ &  GDS\_J1058311-604113 &              $\dots$ \\
AST68688 & $\dots$ &  GDS\_J1058486-611955 &              $\dots$ \\
AST68860 &  312484 &               $\dots$ & OGLE-GD-ECL-05956    \\
AST69265 & $\dots$ &  GDS\_J1058494-610230 &              $\dots$ \\
AST69401 & $\dots$ &  GDS\_J1059213-624329 &              $\dots$ \\
AST69676 & $\dots$ &  GDS\_J1059076-620141 &              $\dots$ \\
AST69995 &  312533 &               $\dots$ & OGLE-GD-ECL-06005    \\
AST70007 & $\dots$ &  GDS\_J1059186-622309 &              $\dots$ \\
AST70054 &  312497 &               $\dots$ & OGLE-GD-ECL-05969    \\
AST70793 & $\dots$ &  GDS\_J1059265-623200 &              $\dots$ \\
AST71022 & $\dots$ &  GDS\_J1058552-605136 &              $\dots$ \\
AST71576 &  312543 &               $\dots$ & OGLE-GD-ECL-06015    \\
AST72278 & $\dots$ &  GDS\_J1058549-601041 &              $\dots$ \\
AST73146 &  312755 &               $\dots$ & OGLE-GD-ECL-06231    \\
AST73648 &    5860 &               $\dots$ & CF Car               \\
AST74334 &    6195 &               $\dots$ & V0442 Car            \\
AST74404 &  312671 &  GDS\_J1059222-604659 & OGLE-GD-ECL-06146    \\
AST74606 &  312741 &               $\dots$ & OGLE-GD-ECL-06217    \\
AST74652 &  312651 &               $\dots$ & OGLE-GD-ECL-06125    \\
AST75081 & $\dots$ &  GDS\_J1059281-605335 &              $\dots$ \\
AST75327 & $\dots$ &  GDS\_J1059141-601331 &              $\dots$ \\
AST75921 &  312765 &               $\dots$ & OGLE-GD-ECL-06242    \\
AST76237 &  312773 &               $\dots$ & OGLE-GD-ECL-06250    \\
AST76422 & $\dots$ &  GDS\_J1059386-611535 &              $\dots$ \\
AST76867 & $\dots$ &  GDS\_J1059356-604139 &              $\dots$ \\
AST77152 &  312841 &  GDS\_J1059546-613515 & OGLE-GD-ECL-06321    \\
AST77247 &  312848 &               $\dots$ & OGLE-GD-ECL-06328    \\
AST77535 & $\dots$ &  GDS\_J1100214-624746 &              $\dots$ \\
AST77614 & $\dots$ &  GDS\_J1059551-612837 &              $\dots$ \\
AST78181 & $\dots$ &  GDS\_J1059356-601639 &              $\dots$ \\
AST78295 & $\dots$ &  GDS\_J1100262-625744 &              $\dots$ \\
AST79966 & $\dots$ &  GDS\_J1059469-601556 &              $\dots$ \\
AST80203 & $\dots$ &  GDS\_J1100036-611357 &              $\dots$ \\
AST80312 & $\dots$ &  GDS\_J1100081-611337 &              $\dots$ \\
AST81880 &  312869 &               $\dots$ & OGLE-GD-ECL-06350    \\
AST82902 & $\dots$ &  GDS\_J1100156-603805 &              $\dots$ \\
AST83717 &    5929 &  GDS\_J1100256-605935 & FM Car               \\
AST84036 & $\dots$ &  GDS\_J1101041-623629 &              $\dots$ \\
AST84689 & $\dots$ &  GDS\_J1100213-601815 &              $\dots$ \\
AST84695 & $\dots$ &  GDS\_J1100568-620445 &              $\dots$ \\
AST85302 & $\dots$ &  GDS\_J1100228-600536 &              $\dots$ \\
AST85912 & $\dots$ &  GDS\_J1100242-600616 &              $\dots$ \\
AST85934 & $\dots$ &  GDS\_J1100453-605746 &              $\dots$ \\
AST86847 &  312949 &               $\dots$ & OGLE-GD-ECL-06431    \\
AST86951 &  312942 &               $\dots$ & OGLE-GD-ECL-06423    \\
AST87214 &  312932 &  GDS\_J1101075-613716 & OGLE-GD-ECL-06413    \\
AST87507 & $\dots$ &  GDS\_J1101127-614518 &              $\dots$ \\
AST87765 &  312943 &               $\dots$ & OGLE-GD-ECL-06424    \\
AST88839 &  312963 &  GDS\_J1101257-615317 & OGLE-GD-ECL-06445    \\
AST88908 & $\dots$ &  GDS\_J1101029-610110 &              $\dots$ \\
AST89238 &  312970 &               $\dots$ & OGLE-GD-ECL-06452    \\
AST90020 & $\dots$ &  GDS\_J1101545-624423 &              $\dots$ \\
AST90095 & $\dots$ &  GDS\_J1101342-621237 &              $\dots$ \\
AST90183 & $\dots$ &  GDS\_J1101134-605947 &              $\dots$ \\
AST90571 & $\dots$ &  GDS\_J1101441-620947 &              $\dots$ \\
AST91731 & $\dots$ &  GDS\_J1101588-622103 &              $\dots$ \\
AST91976 & $\dots$ &  GDS\_J1101339-611052 &              $\dots$ \\
AST92177 & $\dots$ &  GDS\_J1101385-612058 &              $\dots$ \\
AST92381 & $\dots$ &  GDS\_J1101365-611253 &              $\dots$ \\
AST92905 &  313020 &  GDS\_J1102025-620530 & OGLE-GD-ECL-06503    \\
AST93211 &    5930 &  GDS\_J1101143-600659 & FN Car               \\
AST93814 &    5963 &  GDS\_J1101276-602629 & HI Car               \\
AST94842 & $\dots$ &  GDS\_J1101476-605156 &              $\dots$ \\
AST95037 &  313050 &               $\dots$ & OGLE-GD-ECL-06533    \\
AST95098 & $\dots$ &  GDS\_J1102306-623242 &              $\dots$ \\
AST96345 & $\dots$ &  GDS\_J1101525-604511 &              $\dots$ 
\enddata
\end{deluxetable}

\clearpage
\end{document}